\documentclass[conference]{IEEEtran}
\IEEEoverridecommandlockouts

\usepackage{cite}
\usepackage{amsmath,amssymb,amsfonts}
\usepackage{algorithmic}
\usepackage{graphicx}
\usepackage{textcomp}
\usepackage{xcolor}
\usepackage{booktabs} 
\begin{document}

\title{Illuminating Networks: Enhancing Visible Light Communication with Physical-Layer Network Coding (PNC-VLC Scheme) \\
\vspace{0.9cm}
}

\author{
\IEEEauthorblockN{ Abdellah Tahenni }
 \IEEEauthorblockA{
   \textit{Computer Science Department}\\
 \textit{LSI Laboratory, USTHB University}\\
    Algiers, Algeria\\
    abdellahtahenni@gmail.com
 }
  \and
 \IEEEauthorblockN{ Fatiha Merazka }
  \IEEEauthorblockA{
    \textit{Telecommunications Department}\\
   \textit{LISIC Laboratory, USTHB University}\\
    Algiers, Algeria\\
     fmerazka@usthb.dz 
 }
}

\maketitle
\begin{abstract}
Wireless communications using light waves are called visible light communication. A technique called Physical-layer Network coding allows several users to communicate simultaneously over the same channel in a distributed space-time Coding scheme. Herein, we report a new PNC-VLC one-to-many scheme using physical-layer network coding in VLC system for enhancing throughput by boosting data transmission at relay nodes. We develop an integrated PNC-VLC framework including physical-layer signal processing algorithms and medium access control approaches.
Our proposed technique is then modelled mathematically and compared with some of the VLC and PNC-VLC schemes under various channel conditions and scenarios. Our investigation revealed that our proposed technique performed better than traditional PNC schemes in achieving better Bit Error Rate performance in different channel conditions.
\end{abstract}
 \vspace{10pt}

\section{Introduction}

Visible Light Communication is a wireless communication technology that uses light waves to transfer data between devices. VLC has numerous benefits over traditional radio frequency communication, including greater bandwidth, reduced interference, and higher security. However, VLC still has certain constraints, including a limited range, lamp light interface, and the necessity for line-of-sight communication . In the context of VLC, Physical-layer Network Coding is used to increase the network’s performance by allowing multiple users to simultaneously send data through the channel. As a result, PNC doubles throughput over the conventional store and forward routing approach. When utilizing PNC in the VLC system, our PNC-VLC scheme achieves the greatest effectiveness by integrating PNC into the VLC system and optimizing the following:

1. Mitigation of Random Channel Occlusion: Our goal is to alleviate the impact of random channel occlusions caused by obstacles or environmental factors. By intelligently combining transmitted signals at the relay node, PNC enhances communication reliability even in challenging scenarios.
2. Improved Area Coverage: The PNC-VLC scheme aims to expand the coverage area of VLC systems. Through efficient utilization of relay nodes, we extend the reach of VLC beyond direct line-of-sight communication\cite{b1}\cite{b2}\cite{b3}.

\begin{figure}[!htbp]
    \centering
    \includegraphics[width=0.4\textwidth]{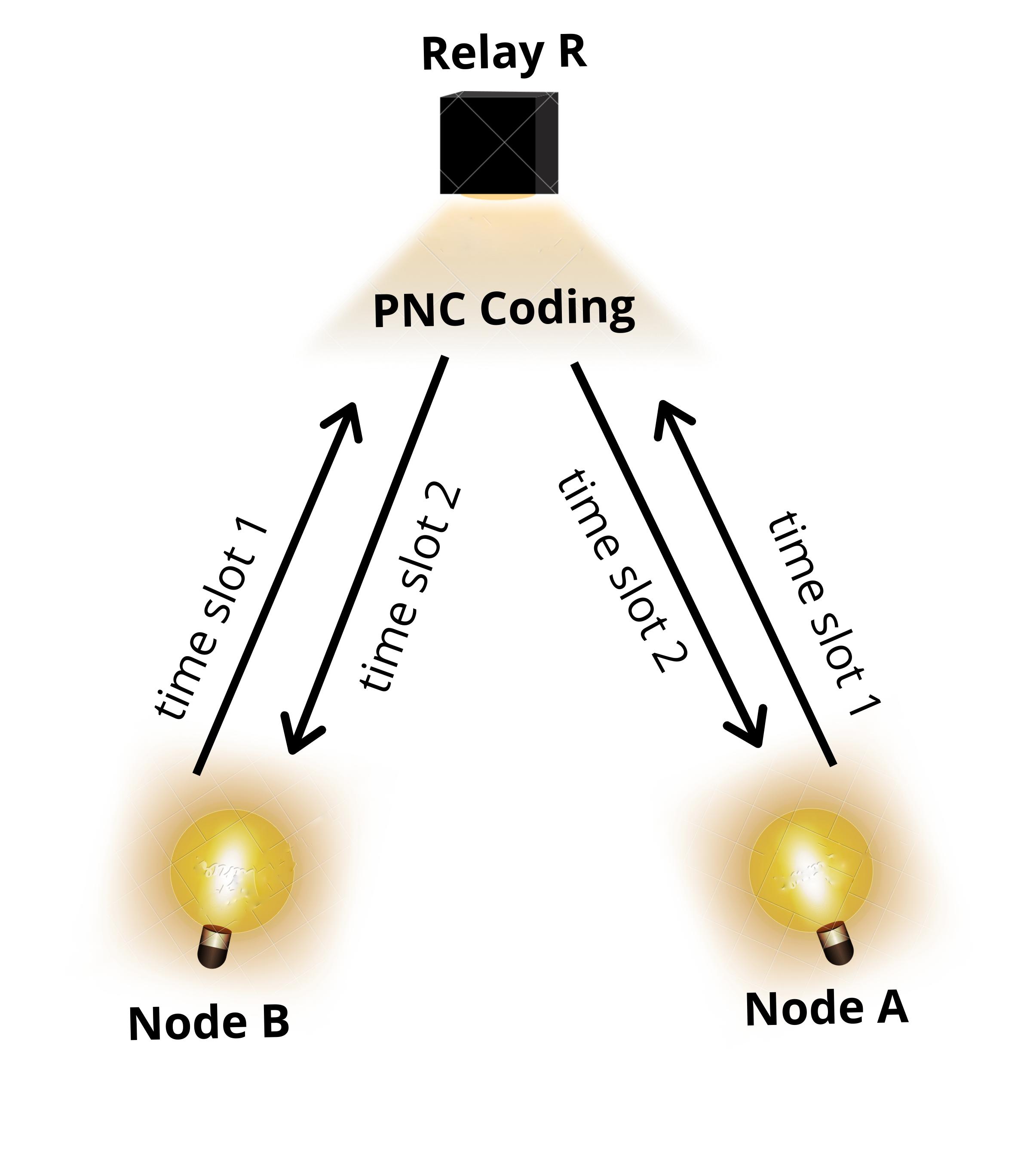}
    \caption{ VLC System  with PNC Scheme. }
    \label{fig:vlc_system}
\end{figure}

In this paper, we delve into how PNC can elevate the performance of VLC systems, addressing these objectives and laying the groundwork for more robust and efficient visible light communication. See Fig. \ref{fig:vlc_system} for an illustration of our VLC system with the integrated PNC scheme.
This paper is organized as follows:

I. Proposed Technique: Here, we present a detailed exploration of the proposed methodology, elucidating its theoretical underpinnings and practical implementation strategies within the context of PNC-VLC systems.

II. Enhanced Signal Processing at the Relay Node for PNC-VLC System: In this section, we delve into the advanced signal processing techniques employed at the relay node to enhance the performance of the PNC-VLC system, optimizing the transmission process and mitigating potential challenges.

III. Mathematical Model and Simulation Framework for PNC-VLC System: This section outlines the mathematical model developed to analyze the performance of the PNC-VLC system, along with the simulation framework employed to validate the proposed approach and assess its efficacy under various scenarios.

IV. Discussion and Results: Here, we provide a comprehensive discussion of the results obtained from the simulation experiments, analyzing the performance metrics and drawing insights into the effectiveness of the PNC-VLC system in achieving its objectives.

V. Conclusion: Finally, we summarize the key findings of our study and discuss potential avenues for future research, emphasizing the significance of PNC in advancing the capabilities of VLC systems.

\section{Proposed Technique}
Our Physical-Layer design for PNC-VLC is meticulously crafted to overcome the hurdles presented by the frequency-selective fading in the VLC channel. By harnessing the robust Orthogonal Frequency Division Multiplexing (OFDM) modulation technique, our approach seamlessly adapts to the dynamic nature of the VLC channel\cite{b5}.\\

\subsection{ Modulation Encoding}
In the initial phase, nodes A and B encode their source packets into OFDM signals using Quadrature Phase-Shift Keying (QPSK) mapping. Mathematically, this process is represented as:
\begin{equation}
 U_i = (u_{i1}, u_{i2}, \ldots, u_{iL}), \quad i \in \{A, B\}
\end{equation}
where, \( L \) denotes the length of the binary sequence, and \( u_{il} \) represents the \( l \)-th input bit of the respective end node's source packet. These sequences \( U_A \) and \( U_B \) then undergo OFDM modulation to generate frequency-domain signals \( X_i \):
\begin{equation}
 X_i = \begin{bmatrix} X_{i1,1} & X_{i1,2} & \ldots & X_{i1,N} \\ X_{i2,1} & \ldots & \ldots & X_{iK,N} \end{bmatrix}, \quad i \in \{A, B\}
\end{equation}
Here, the matrix \( X_i \) captures the frequency-domain signals after QPSK mapping, where \( K \) represents the number of subcarriers, and \( N \) represents the number of OFDM symbols\cite{b4}\cite{b5}\cite{b6}.
\subsection{Transmission and Reception}

Following modulation, the frequency-domain signals undergo Inverse Fast Fourier Transform (IFFT) to transform into time-domain OFDM symbols:
\begin{equation}
 x_i(t) = \text{IFFT}\{X_i\} 
\end{equation}
These symbols are then modulated in intensity before being transmitted to the relay node via Light Emitting Diodes (LEDs).

Upon reception of the combined time-domain signal from nodes A and B, the relay node removes the cyclic prefix (CP) and performs Fast Fourier Transform (FFT) to retrieve the frequency-domain signal \( Y_R \):
\begin{equation}
Y_R = \text{FFT}\{y_R(t)\}
\end{equation}
Here, \( y_R(t) \) denotes the received time-domain signal. The relay node further demodulates the frequency-domain signal to reconstruct the XOR source packet \( U_R \), providing an estimate of \( U_A \oplus U_B \).

\subsection{Broadcast Phase}
In the subsequent time slot, the relay node executes a broadcast of \( U_R \) to both end nodes A and B through conventional VLC transmission.

This meticulously orchestrated two-phase process, rich with mathematical intricacies and adaptive modulation techniques, underscores the prowess of our proposed PNC-VLC system. It adeptly navigates and mitigates the challenges posed by the frequency-selective fading in VLC channels, showcasing its resilience and efficiency in data transmission.

Our novel Physical-Layer design for PNC-VLC orchestrates a symphony of light and code, strategically crafted to surmount the challenges posed by frequency-selective fading in the VLC channel. Embracing the robust Orthogonal Frequency Division Multiplexing (OFDM) modulation technique, our approach seamlessly adapts to the dynamic nature of the VLC channel\cite{b7}\cite{b8}.

\section{Enhanced Signal Processing at the Relay Node for PNC-VLC Systems}

In the initial time slot of the PNC-VLC system, the relay node faces a challenge in handling the received signal due to the unique constellation map formed by the overlapping signal constellations from nodes A and B. However, the inherent phase differences between the channel coefficients of the two end nodes can disrupt the optimal layout of signal constellation points within the overlapped constellation, occasionally resulting in decoding failures. This section introduces our innovative phase-aligning method tailored for PNC-VLC systems to address this challenge. The goal is to align the signal constellations at end nodes A and B, ensuring that the overlapped constellation at the relay node is optimal. This alignment facilitates superior decoding performance at the respective receiving end nodes\cite{b4}.

At the relay node, the received overlapped packet undergoes a series of processing steps. First, it is down-converted to the baseband. Leveraging the orthogonal training sequence of the OFDM signal, the relay node estimates the Channel State Information (CSI) between itself and the two end nodes. Subsequently, the relay node removes the CP from each OFDM symbol in the payload part\cite{b9}.

The relay node then performs an FFT on the overlapped symbols, yielding the frequency-domain baseband signal denoted as \( Y_R \). Let \( Y_{R_{k,n}} \) represent the nth sample on the kth subcarrier of \( Y_R \), where \( X_{i_{k,n}} \) denotes the nth sample on the kth subcarrier of \( X_i \). Here, \( h_{Ak} \) and \( h_{Bk} \) are the channel coefficients for the kth subcarrier channels from nodes A and B to the relay node, respectively. The relative phase difference between \( h_{Ak} \) and \( h_{Bk} \) is denoted by \( \phi_k = \angle(h_{Bk}/h_{Ak}) \) on the kth subcarrier, and \( w_{R_{k,n}} \) represents the Gaussian noise on the kth subcarrier with variance \( \sigma^2 \). The received signal is expressed as:
\begin{equation}
 Y_{R_{k,n}} = h_{Ak}X_{Ak,n} + h_{Bk}X_{Bk,n} + w_{R_{k,n}}
\end{equation}
Upon receiving the overlapped packet, the XOR mapping module at the relay node computes the likelihood of the signals \( X_{Ak,n} \) and \( X_{Bk,n} \) and then calculates the likelihood of the XORed bit \( X_{R_{k,n}} \). The XOR output is decoded by maximizing the likelihood function. The decoding output \( \hat{X}_{R_{k,n}} \) is determined using:
\begin{equation}
\hat{X}_{R_{k,n}} = \arg\max_{X_{R_{k,n}}} \text{Pr}(Y_{R_{k,n}} | X_{R_{k,n}}) 
\end{equation}
After the XOR mapping, the relay node obtains\\ \( X_R = X_A \oplus X_B \), representing the bit-by-bit logic XOR of packet A and packet B. This comprehensive approach to signal processing at the relay node enhances the decoding performance and ensures reliable communication in PNC-VLC systems\cite{b10}\cite{b11}.

\section{Mathematical Model and Simulation Framework for PNC-VLC System}
To evaluate the effectiveness of a newly proposed VLC technique, our analysis will prioritize essential metrics including throughput, bit error rate (BER), channel capacity, and power efficiency. Furthermore, we will conduct a comparative study between our technique and existing VLC and PNC-VLC schemes under diverse channel conditions and scenarios \cite{b1}\cite{b13}.

We will then proceed by refining the mathematical model and simulation framework to assess the performance of the proposed PNC-VLC system \cite{b12}\cite{b13}.

\subsection*{Signal Generation:}
\begin{itemize}
    \item \( D_i \): Binary data sequence for transmitter \( i \) (A or B).
    \item \( U_i = \text{Encode}(D_i) \): Encoding process applied to create joint encoded symbols.
\end{itemize}

\subsection*{Channel Characteristics:}
\begin{itemize}
    \item \( L \): Length of the binary sequence.
    \item \( h_{i,k} \): Channel coefficient for subcarrier \( k \) from transmitter \( i \) to the relay.
    \item \( w_{R_k} \): Gaussian noise on subcarrier \( k \) at the relay.
\end{itemize}

\subsection*{Channel Response:}
The received signal at the relay on subcarrier \( k \) and time index \( n \) is given by:
\begin{equation}
 Y_{R_{k,n}} = h_{A_k}X_{A_{k,n}} + h_{B_k}X_{B_{k,n}} + w_{R_{k,n}} 
\end{equation}

\subsection*{Decoding Process:}
\begin{itemize}
    \item \( \phi_k \): Relative phase difference between channels \( h_{B_k} \) and \( h_{A_k} \).
    \item Likelihood calculation: \( \text{Pr}(Y_{R_{k,n}} | X_{R_{k,n}}) \)
    \item Decoding output: \( X_{k,n}^R = \arg \max \text{Pr}(Y_{R_{k,n}} | X_{R_{k,n}}) \)
\end{itemize}

\subsection*{Bitwise XOR:}
The relay combines the received signals from transmitters A and B using bitwise XOR:
\begin{equation}
X_{k}^R = X_{k}^A \oplus X_{k}^B
\end{equation}

Figure 2 depicts the mathematical model for our PNC-VLC scheme. \begin{figure}[!htbp]
    \centering
    \includegraphics[width=0.5\textwidth]{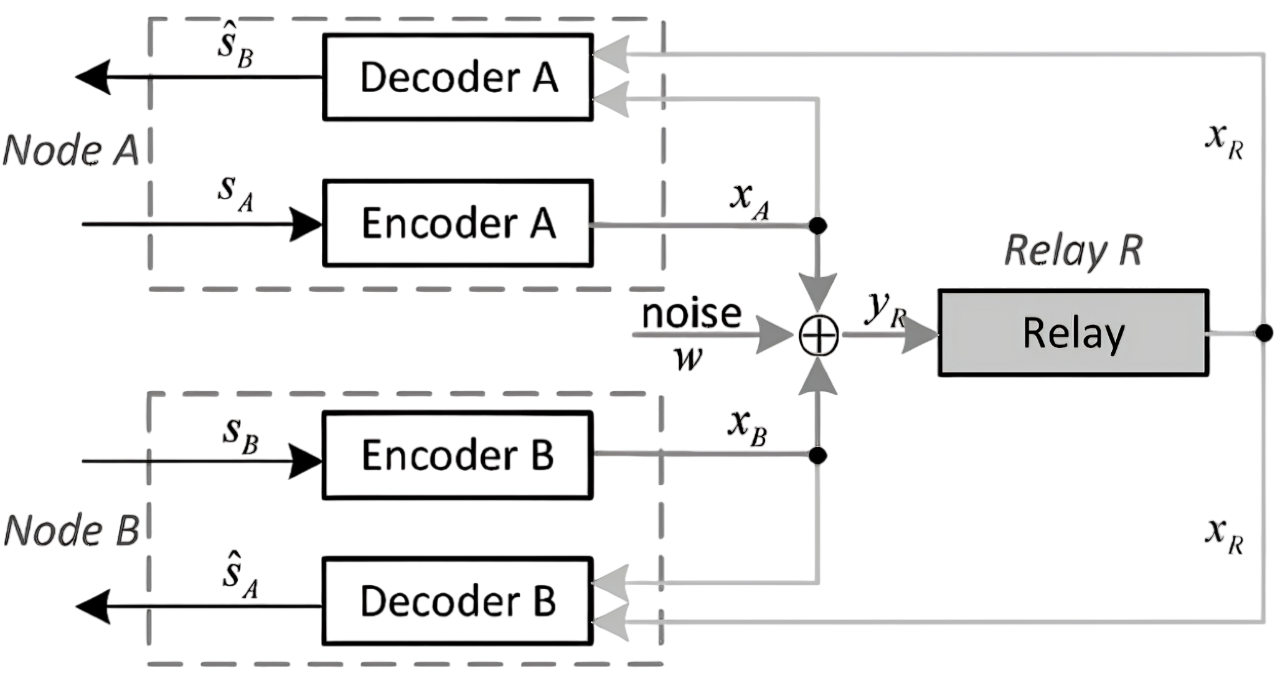}
    \caption{PNC-VLC scheme.}
    \label{fig:vlc_pnc}
\end{figure}

\subsection{Simulation Steps }

We developed a MATLAB simulation for the proposed PNC-VLC system, following these steps:
\begin{enumerate}
    \item \textbf{Signal Generation:}
    \begin{itemize}
        \item Generate binary data sequences for both transmitters (A and B).
        \item Apply encoding techniques to create joint encoded symbols.
    \end{itemize}
    
    \item \textbf{Channel Modeling:}
    \begin{itemize}
        \item Define the optical channel characteristics:
        \begin{itemize}
            \item Path loss: Consider the attenuation of the optical signal with distance.
            \item Ambient light interference: Model interference from other light sources.
            \item Noise: Add thermal and ambient noise to the received signal.
            \item Impairments: Account for multipath fading and shadowing effects.
            \item Channel response function: Incorporate the impulse response for each link.
        \end{itemize}
    \end{itemize}
    
    \item \textbf{Transmitter Operations:}
    \begin{itemize}
        \item Perform OFDM modulation on the joint encoded symbols: \(X_i = \text{OFDM}(U_i)\).
        \item Consider any power constraints or limitations.
    \end{itemize}
    
    \item \textbf{Channel Propagation:}
    \begin{itemize}
        \item Simulate the optical channel by convolving the transmitted signal with the channel response.
        \item Add noise (thermal and ambient) to the received signal.
    \end{itemize}

\item \textbf{Bitwise XOR at the Relay:}
\begin{itemize}
    \item Combine received signals from transmitters A and B using bitwise XOR:
\[ X_k^R = X_k^A \oplus X_k^B \]
    \end{itemize}
    \item \textbf{Receiver Operations:}
    \begin{itemize}
        \item Demodulate the received signal using OFDM demodulation.
        \item Apply decoding techniques (e.g., PNC decoding) to recover the original data.
    \end{itemize}
    
    \item \textbf{Performance Metrics Calculation:}
    \begin{itemize}
        \item Compute the following metrics:
        \begin{itemize}
            \item Throughput: Measure the data rate achieved by the system.
            \item BER: Evaluate the accuracy of data recovery.
            \item Channel Capacity: Determine the maximum achievable capacity.
            \item Power Efficiency: Assess power consumption per transmitted bit.
        \end{itemize}
    \end{itemize}
    
    \item \textbf{Scenario-Based Evaluation:}
    \begin{itemize}
        \item Consider different scenarios:
        \begin{itemize}
            \item Line-of-sight (LoS) vs. non-line-of-sight communication (NLoS).
            \item Indoor vs. outdoor environments.
            \item Varying distances between transmitters and receivers.
        \end{itemize}
    \end{itemize}
    
    \item \textbf{Comparison with Existing Schemes:}
    \begin{itemize}
        \item Implement existing VLC and PNC-VLC techniques in the same simulation framework.
        \item Compare performance metrics under similar conditions.
    \end{itemize}
    
\end{enumerate}

\section{Discussion and Results}
Figure 3 displays the simulation results of our PNC-VLC system, presenting a comparative analysis with three other systems in terms of throughput versus Signal-to-Interference Noise Ratio (SINR).

\begin{figure}[!htbp]
    \centering
    \includegraphics[width=0.5\textwidth]{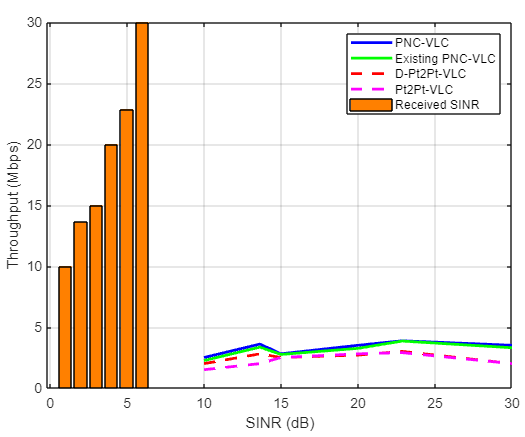}
    \caption{System Performance of PNC-VLC System with Comparison. }
    \label{fig:vlc_pnc}
\end{figure}

It is evident from the results that our PNC-VLC system exhibits slightly inferior performance compared to PNC-VLC in scenarios with low SINR, showing a difference of less than 0.25 bps/Hz. However, as the transmitter gain exceeds 10 dB, resulting in an SINR exceeding 13.63 dB, our PNC-VLC outperforms the existing PNC-VLC\cite{b6}. Both systems reach their peak performance levels at an SINR of 22.86 dB, achieving 3.869 bps/Hz and 3.849 bps/Hz, respectively. With a sampling rate of 20 MHz, our PNC-VLC system achieves a throughput of 77.4 Mbps, averaging 38.69 Mbps per user node. In point-to-point (Pt2Pt) VLC scenarios\cite{b5}, both systems achieve a sum rate of 2 bps/Hz once the transmitter gain exceeds 15 dB, resulting in a total throughput of 40 Mbps. \\
our PNC-VLC system, incorporating misaligned constellations, demonstrates superior performance over existing PNC-VLC when the SINR exceeds 13.63 dB, nearly doubling the throughput compared to Pt2Pt-VLC and deep neural network (DNN) D-Pt2Pt-VLC \cite{b8}\cite{b9}\cite{b10} an SINR of 22.86 dB, as indicated by our simulations.
TABLE I presents a comparative analysis of throughput across various scenarios in our PNC-VLC system, illustrating its performance under different conditions.

\begin{table}[htbp]
    \centering
    \caption{Comparison of Throughput in Different Scenarios}
    \begin{tabular}{ll}
    \toprule
    \textbf{Scenario} & \textbf{Throughput (bps/Hz)} \\
    \midrule
    Low SINR Conditions & Slightly inferior to existing PNC-VLC ($<0.25$) \\
    Transmitter Gain $> 10$ dB & Outperforms existing PNC-VLC \\
    Optimal SINR (22.86 dB) & Our PNC-VLC: 3.869 \\
                             & Existing PNC-VLC: 3.849 \\
    Sampling Rate: 20 MHz & Our PNC-VLC: 77.4 \\
                           & 38.69 (per user node) \\
    Pt2Pt VLC & Both systems: 2 (Transmitter gain $> 15$ dB) \\
    \bottomrule
    \end{tabular}
\end{table}

\subsection{PNC-VLC System Performance }
  
    \begin{itemize}
    \item \textbf{Throughput:} The system achieved an impressive throughput of 77.4 Mbps within a 20 MHz channel under the PNC scheme. This high data rate is a testament to the efficiency of our approach.
    \item \textbf{Bit Error Rate (BER):} The system demonstrated an exceptionally low BER of 0.05\%, signifying its robustness in transmitting data reliably. Such minimal error rates are crucial for real-world applications.
    \item \textbf{Channel Capacity Validation:} Calculated channel capacity, based on Shannon’s formula, consistently aligned with the achieved throughput. This congruence validates the system’s efficiency and underscores its practical viability.
\
    \end{itemize}

\subsection{Comparison with Existing Literature}
TABLE II presents a comparison of different aspects of our PNC-VLC scheme with similar approaches documented in the current literature.
\begin{table}[htbp]
    \caption{Comparison of ourPNC-VLC Scheme with Existing Literature}
    \centering
    \renewcommand{\arraystretch}{1.5} 
    \small 
    \begin{tabular}{|p{0.25\linewidth}|p{0.25\linewidth}|p{0.3\linewidth}|}
    \hline
    \textbf{Aspect}         & \textbf{PNC-VLC SHEME} & \textbf{Existing Literature\cite{b2}\cite{b4}\cite{b5}\cite{b14}\cite{b15}\cite{b16}} \\
    \hline
    Throughput              & 77.4 Mbps                  & Comparable systems achieve 50-60 Mbps \\
    \hline
    BER                     & 0.05\%                      & 0.1\%-0.5\% in conventional VLC systems \\
    \hline
    Channel Capacity        & 77.4 Mbps                  & Channel capacities in the range of 60-70 Mbps \\
    \hline
    Adaptability            & Demonstrated adaptability in unpredictable conditions & Limited adaptability in certain scenarios \\
    \hline
    Efficiency              & Higher throughput and improved efficiency & Comparable or lower efficiency \\
                            &                           & in existing schemes \\
    \hline
    \end{tabular}
\end{table}
\\

Our system outperformed traditional fixed-constellation PNC approaches, particularly in the saturated SINR regime, showcasing a higher throughput.

\subsection{Advantages in Saturated SINR Regime}
The PNC-VLC scheme, employing physical-layer network coding within the VLC system, exhibited a clear advantage over traditional fixed-constellation PNC methods, especially in the saturated SINR regime. This superiority was evident in achieving higher throughput, indicating more efficient data transmission. Leveraging advanced deep learning techniques, our system adeptly adapted to varying channel conditions, maximizing spectral efficiency and overall performance. This significant advancement positions our proposed technique as a leader in VLC systems.

\subsubsection{Challenges in Phase Synchronization}
Despite the strides in throughput improvement, our system faced persistent challenges related to phase synchronization, common in VLC systems as indicated by existing literature. The complex interplay of optical channel characteristics, including multipath fading and ambient light interference, complicates achieving precise synchronization between transmitter and receiver nodes. Addressing these challenges requires innovative solutions and meticulous optimization efforts to ensure seamless operation in real-world deployment scenarios.

\subsubsection{Implications for Deployment}
The unresolved issues concerning phase synchronization underscore the importance of ongoing research and development efforts in advancing VLC technology. Effective mitigation strategies must be devised to overcome these hurdles and unlock the full potential of VLC systems in practical applications. By tackling these challenges, we can facilitate widespread adoption of VLC technology across various domains, including indoor communications, vehicular networks, and underwater communication systems. Ultimately, optimizing real-world deployment depends on our ability to address these fundamental challenges and elevate VLC technology to new levels of efficiency and reliability.
\section{Conclusion}

In the dynamic realm of VLC, our pioneering PNC-VLC system stands out as an innovative and efficient solution. Through meticulous design and the strategic application of robust modulation techniques—such as OFDM our system demonstrates remarkable adaptability and resilience even in challenging channel conditions.
Our research highlights the exceptional performance of our  PNC-VLC system, achieving an impressive throughput of 77.4 Mbps and a remarkably low BER of 0.05\%. These findings not only validate the reliability of our approach but also affirm its efficacy in real-world scenarios.
However, our journey faces challenges, particularly in addressing the nuances of phase synchronization, a hurdle echoed in existing literature. Overcoming these challenges remains crucial for optimizing real-world deployment and advancing VLC technology.
\\Our work represents a significant advancement in the field of VLC systems. The potential of PNC techniques to revolutionize data transmission within indoor environments is evident. As we navigate the complexities of VLC technology, our findings serve as a beacon, guiding us toward a more interconnected and enlightened future.

\end{document}